\documentclass[aps,amsmath,amssymb,reprint,superscriptaddress]{revtex4-1}

\newcommand{\themax}{$\rho_{\textrm{max}}^{\textrm{THE}}$~}
\newcommand{\ahe}{$\rho^{\textrm{AHE}}$~}
\newcommand{\mpsa}{Mn\textsubscript{2}PtSn~}
\newcommand{\mpsb}{Mn\textsubscript{1.5}PtSn~}

\usepackage{graphicx}% Include figure files
\usepackage{dcolumn}% Align table columns on decimal point
\usepackage[utf8]{inputenc}
\usepackage{natbib}
\usepackage{siunitx}
\usepackage[hyperindex,breaklinks]{hyperref}
\usepackage{verbatim}

\begin{document}

\title{Topological {Hall} effect in thin films of {Mn$_{1.5}$PtSn} }

\author{Peter Swekis}
    \affiliation{Max-Planck Institute for Chemical Physics of Solids, 01187 Dresden, Germany}
    \affiliation{Institut f{\"u}r Festk{\"o}rper- und Materialphysik, Technische Universit{\"a}t Dresden, 01062 Dresden, Germany}
\author{Anastasios Markou}
    \email{Anastasios.Markou@cpfs.mpg.de}
    \affiliation{Max-Planck Institute for Chemical Physics of Solids, 01187 Dresden, Germany}
\author{Dominik Kriegner}
    \affiliation{Max-Planck Institute for Chemical Physics of Solids, 01187 Dresden, Germany}
\author{Jacob Gayles}
    \affiliation{Max-Planck Institute for Chemical Physics of Solids, 01187 Dresden, Germany}
\author{Richard Schlitz}
    \affiliation{Institut f{\"u}r Festk{\"o}rper- und Materialphysik, Technische Universit{\"a}t Dresden, 01062 Dresden, Germany}
    \affiliation{Center for Transport and Devices of Emergent Materials, Technische Universit{\"a}t Dresden, 01062 Dresden, Germany}
\author{Walter Schnelle}
    \affiliation{Max-Planck Institute for Chemical Physics of Solids, 01187 Dresden, Germany}
\author{Sebastian T. B. Goennenwein}
    \affiliation{Institut f{\"u}r Festk{\"o}rper- und Materialphysik, Technische Universit{\"a}t Dresden, 01062 Dresden, Germany}
    \affiliation{Center for Transport and Devices of Emergent Materials, Technische Universit{\"a}t Dresden, 01062 Dresden, Germany}
\author{Claudia Felser}
    \email{Claudia.Felser@cpfs.mpg.de}
    \affiliation{Max-Planck Institute for Chemical Physics of Solids, 01187 Dresden, Germany}

\date{\today}

\begin{abstract}
Spin chirality in metallic materials with non-coplanar magnetic order can give rise to a Berry phase induced topological Hall effect.
Here, we report the observation of a large topological Hall effect in high-quality films of Mn$_{1.5}$PtSn that were grown by means of magnetron sputtering on MgO(001). The topological Hall resistivity is present up to $\mu_{0}H \approx 4~$T below the spin reorientation transition temperature, $T_{s}=185$~K. We find, that the maximum topological Hall resistivity is of comparable magnitude as the anomalous Hall resistivity. Owing to the size, the topological Hall effect is directly evident prior to the customarily performed subtraction of magnetometry data. Further, we underline the robustness of the topological Hall effect in Mn\textsubscript{2-x}PtSn by extracting the effect for multiple stoichiometries (x~=~0.5, 0.25, 0.1) and film thicknesses (t = 104, 52, 35~nm) with maximum topological Hall resistivities between $0.76~\mu\Omega$cm and $1.55~\mu\Omega$cm at 150~K.
\end{abstract}

\maketitle

\section{\label{sec:level1}Introduction}
Topological magnetic structures have become of great interest recently, attributed to the emergent transport phenomena associated with the magnetic texture~\cite{Fert2013}. One of these phenomena is the transverse Hall current, that arises from the interplay of magnetic order and intrinsic band structure or scattering~\cite{Nagaosa2010}.  Experimentally, the measured Hall resistivity can be separated into the ordinary Hall effect (OHE)~\cite{Hall1879} dependent on the external field ($H$) and the anomalous Hall effect (AHE) which scales with the saturation magnetization. The modern understanding of the AHE ascribes the effect to scattering mechanisms~\cite{Karplus1954,Smit1958,Berger1970} and the intrinsic momentum space Berry curvature~\cite{Fang2003}. However, recently an additional Hall-type contribution was proposed that scales neither with the magnetization ($M$) nor with the externally applied field, termed the topological Hall effect (THE)~\cite{Ye1999,Bruno2004}. This THE has been proposed to originate from a finite scalar spin chirality~\cite{Taguchi2001}, skyrmions~\cite{Ye1999}, and Weyl points~\cite{Kuebler2014}. The prior two are connected through the magnetic texture and the latter is connected to the momentum space dispersion. Here, we focus on the magnetic texture induced THE which has become of great interest in Heusler compounds due to their tunability~\cite{Manna2018,Wollmann2017}.

There are two limiting cases for the stabilization of magnetic textures: the scalar spin chirality and the skyrmionic lattice~\cite{Roessler2006}, which originate from a competition of exchange, e.g. Heisenberg and Dzyaloshinki-Moriya interaction~\cite{Dzyaloshinsky1958,Moriya1960}, with anisotropy and external fields. In the limit of discrete spins, there is a finite scalar spin chirality $S_{i}\cdot (S_{j} \times S_{k})$ caused by three non-coplanar spins that subtend a finite cone angle and give rise to the momentum space dependent THE~\cite{Taguchi2001,Suergers2014}. In the adiabatic limit, the spin chirality is taken to be continuous as the integer winding of the real space Berry curvature~\cite{Ye1999,Bruno2004,NeuBauer2009}. As electrons couple to such spin textures, they acquire a finite Berry phase acting as a magnetic field. This in turn results in an additional contribution to the Hall effect~\cite{Berry1984}.

The THE has been observed in a variety of materials including the B20 compounds~\cite{NeuBauer2009,Kanazawa2011,Huang2012}, perovskites~\cite{Nakamura2018,Vistoli2018} and Heusler compounds~\cite{Rana2016,Liu2018,Li2018}. The Heusler compounds are of particular interest, owing to the recent discovery of antiskyrmions in Mn\textsubscript{1.4}Pt\textsubscript{0.9}Pd\textsubscript{0.1}Sn, a new type of topological texture due to the \textit{$D_{2d}$} symmetry~\cite{Nayak2017}. 
The ferrimagnetic Mn\textsubscript{2}YZ (Y being a transition metal and Z a main-group element) inverse Heusler compounds that crystallize in a non-centrosymmetric structure with \textit{$D_{2d}$} symmetry are promising candidates to realize such spin textures through competing interactions of the magnetic sublattices and magnetocrystalline anisotropy caused by tetragonal distortion~\cite{Meshcheriakova2014}. In thin films the presence of geometric constraints can additionally stabilize the desired spin textures in a wider field and temperature range~\cite{Butenko2010}. Recently, the THE was observed in single crystal thin films of Mn\textsubscript{2}RhSn~\cite{Rana2016} as well as in bulk \mpsa\cite{Liu2018} below a spin reorientation transition temperature ($T_{s}$)~\cite{Meshcheriakova2014}, and in polycrystalline \mpsa\cite{Li2018} films for all temperatures below the Curie temperature. Conversely, the work of Jin \textit{et al.} shows no topological Hall signal or $T_{s}$ in epitaxially grown films of \mpsa\cite{Jin2018}.   

In this paper, we focus on \mpsb thin films, with the closest stoichiometry relation to the antiskyrmion compound Mn\textsubscript{1.4}Pt\textsubscript{0.9}Pd\textsubscript{0.1}Sn. We demonstrate the presence of a THE below a spin reorientation transition temperature $T_{s}$ and up to high fields, evident prior to the customarily performed subtraction of magnetometry data. Further, we point out the robustness of the THE in Mn$_{2-x}$PtSn by comparing different compositions and film thicknesses, as well as previously reported results on Mn$_{2}$PtSn films.

%%%%%%%%%%%%%%%%%%%%%%%%%%%%%%%%%%%%%%%%%%%%%%%%%%%%%%%%%%%%%

\section{\label{sec:level2}Experimental Details}
High-quality Mn\textsubscript{2-x}PtSn films were grown on single crystal MgO (001) substrates in a BESTEC UHV magnetron sputtering system. Mn, Pt and Sn were deposited from 2'' targets using DC magnetron co-sputtering. The stoichiometry was controlled by adjusting the power of the magnetrons. The deposition was performed in confocal geometry with a target to substrate distance of 200~mm. Prior to deposition, the chamber was evacuated to a base pressure below $2\times 10\textsuperscript{-8}$~mbar, while during deposition a process gas pressure of $3\times 10~\textsuperscript{-3}$~mbar (Ar, 15~sccm) was maintained. The films were deposited at \SI{350}{\celsius} and post-annealed for 30~min at the same temperature in order to improve the chemical ordering. The annealed films were capped at room temperature with 3~nm Al, in order to prevent oxidization.

The film compositions were confirmed using energy-dispersive x-ray (EDX) microscopy. The film surface topography was analyzed by atomic force microscopy (AFM) on an Asylum Research MFP-3D Origin by Oxford Instruments. Structural characterization was carried out using x-ray diffractometry (XRD) with Cu-K$\alpha$ 1 radiation ($\lambda =1.5406$ \AA) on a PANalytical X'Pert PRO system. The film thickness ($t$) was determined by x-ray reflectivity (XRR) measurements.

Magnetization measurements were performed on a vibrating sample magnetometer (MPMS 3, Quantum Design). In order to infer the magnetization of the films, we subtracted the diamagnetic substrate contribution as well as a low-temperature paramagnetic contribution from the raw data. Here, the paramagnetic contribution can be attributed to impurities in the MgO substrate. The diamagnetic susceptibility ($\chi=-19.066\times 10\textsuperscript{-6}$) of MgO was determined from reference measurements. The paramagnetic contribution was fitted and subtracted from the raw data using the Brillouin function.

Four-probe and five-probe measurements were performed to obtain the resistivity along the longitudinal direction and the Hall resistivity, respectively. Therefore, an in-plane current, $I_{x}=50~\mu\mathrm{A}$, was applied along a film stripe with a width of $W=1.29~\mathrm{mm}$ ($y$-direction). Voltages were recorded simultaneously along the current direction ($V_{x}$), with a lead distance of $L=7.4~\mathrm{mm}$ ($x$-direction), as well as perpendicular to the current direction ($V_{y}$), with a lead distance of $w=0.77~\mathrm{mm}$. The magnetic field was applied along the out-of-plane ($z$) direction (MgO [001]). 
In order to obtain a clean resistivity, $\rho_{xx}$, the raw resistivity, $\rho_{xx}^{\mathrm{raw}}=V_{x}Wt/(LI_{x})$, was symmetrized by averaging $\rho_{xx}^{\mathrm{raw}}$ at positive and negative fields with respect to the field sweep directions. To obtain a clean Hall resistivity, $\rho_{xy}$, the raw Hall resistivity, $\rho_{xy}^{raw}=V_{y}Wt/(wI_{x})$, was antisymmetrized by averaging the difference of $\rho_{xy}^{\mathrm{raw}}$ at positive and negative fields with respect to the field sweep directions.

%%%%%%%%%%%%%%%%%%%%%%%%%%%%%%%%%%%%%%%%%%%%%%%%%%%%%%%%%%%%%

\section{\label{sec:level3}Results and Discussion}
\subsection{\label{sec:level3a}Structural Characterization}
In the following, we discuss the properties of a Mn$_{1.5}$PtSn thin film in detail, since it has the closest stoichiometry relation to the antiskyrmion compound Mn\textsubscript{1.4}Pt\textsubscript{0.9}Pd\textsubscript{0.1}Sn. In Fig. 1(a) we show the x-ray reflectivity, together with AFM analysis, confirming the smoothness of the film with a r.m.s roughness of 0.3~nm in the obtained 5~$\mu$m $\times$ 5~$\mu$m scan. The Kiessig fringes, reaching beyond the measurement range, are further evidence of a high-quality surface as well as a high-quality substrate to film interface. A thickness of 104.7~nm and a roughness of less than 0.5~nm is inferred from XRR fitting. 

\begin{figure}
    \centering
    \includegraphics[width=\columnwidth]{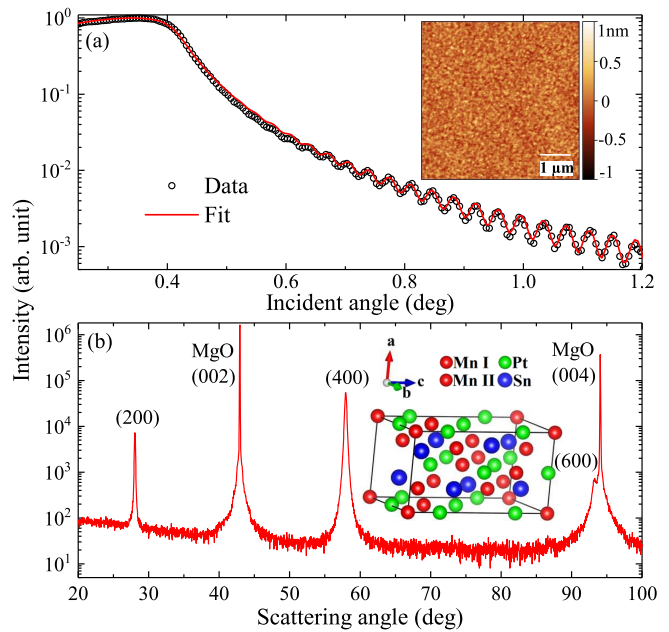}
    \caption{Structural characterization of the 104~nm thick Mn$_{1.5}$PtSn film. (a) XRR pattern with Kiessig fringes including fit. Inset: AFM image. (b) $\omega$ - $2\theta$-scans recorded in out-of-plane geometry showing the (200), (400) and (600) peaks as well as the (002) and (004) substrate Bragg peaks. Inset: tetragonal crystal structure.}
    \label{fig:XRD}
\end{figure}

Furthermore, we use x-ray diffraction radial scans ($\omega-2\theta$) as shown in Fig.~\ref{fig:XRD}(b) and  Fig.~S1 (see the Supplemental Material~\cite{Supp}) to determine the crystal structure of our film. The symmetric radial scans in Fig.~\ref{fig:XRD}(b) confirm epitaxial growth since only the ($h$00) series of Bragg peaks, attributed to the \mpsb film, can be observed. The full-width at half-maximum of the (400) out-of-plane rocking curve of 1.147$^{\circ}$~verifies high crystallinity. Additionally, more than 10 asymmetric Bragg peaks (Fig.~S1) can be indexed using a unit cell similar to bulk Mn\textsubscript{1.4}PtSn~\cite{Nayak2017}. Analogous to the bulk structure, we describe our unit cell by the space group $I\overline{4}2d$ (\#122), which is derived from the inverse tetragonal Heusler structure. This is supported by the observation of a systematic absence of Bragg peaks corresponding to this crystal symmetry (Fig.~S1). By modeling the peak intensities we find that Mn atoms occupy the $4c$ and $8d$ ($x=0.75$) positions, while the Pt and Sn atoms occupy the $8c$ ($z=0.23$) and $8d$ ($x=0.29$) positions, respectively. A detailed analysis of the peak positions shows that the film geometry stabilizes the $c$ axis in the film plane, slightly breaking the equivalence of the $a$ and $b$ parameters, reflected in the lattice parameters $a=6.338$~\AA~$\pm~0.004$~\AA, $b=6.36$~\AA~$\pm~0.01$~\AA~and $c=12.22$~\AA~$\pm~0.03$~\AA.

From the \{112\} pole figure and the comparison of the corresponding azimuthal scan as well as the splitting of high-angle peaks (see Fig.~S1 and S2 in the Supplemental Material~\cite{Supp}) we conclude that two orientations of the $c$ axis, along [$1\bar10$] and [110] of the MgO substrate are present. For the two lattice directions within the film plane, this corresponds to a lattice mismatch of 2.5\% and 6.5\%, respectively.

%%%%%%%%%%%%%%%%%%%%%%%%%%%%%%%%%%%%%%%%%%%%%%%%%%%%%%%%%%%%%

\subsection{\label{sec:level3b}Magnetometry and Magnetotransport Properties}
\begin{figure}
    \centering
    \includegraphics[width=\columnwidth]{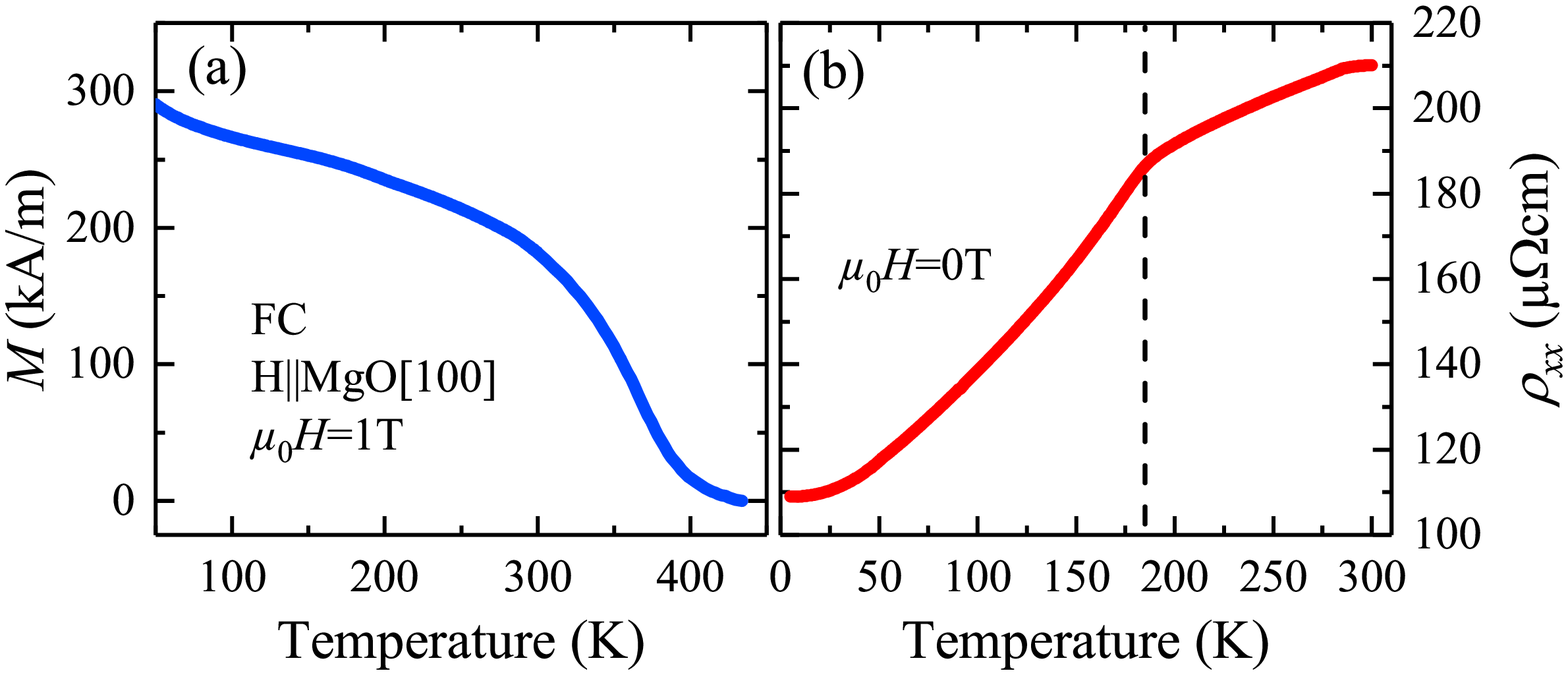}
    \caption{In-plane magnetization (a) and zero field resistivity (b) of the 104~nm thick \mpsb film as a function of temperature. The kink at $T_{s}=185$~K in (b) marks the spin reorientation transition temperature. }
    \label{fig:temp}
\end{figure}

Figure~\ref{fig:temp}(a) depicts the temperature dependence of the magnetization at 1~T with a single transition at 400~K representing the Curie temperature for the 104~nm thick Mn$_{1.5}$PtSn film. A spin reorientation is not clearly evident for this field. Figure~\ref{fig:temp}(b) shows the temperature dependence of the longitudinal resistivity. In analogy to the case of Mn$_{2}$RhSn~\cite{Meshcheriakova2014}, a change in the slope at $T_{s}=185$~K marks a transition from a collinear ($T>T_{s}$) into a non-collinear ($T<T_{s}$) magnetic structure following spin reorientation of one Mn sub-lattice. A similar feature was also observed in related compounds~\cite{Liu2018,Nayak2017,Li2018}.

\begin{figure}
    \centering
    \includegraphics[width=\columnwidth]{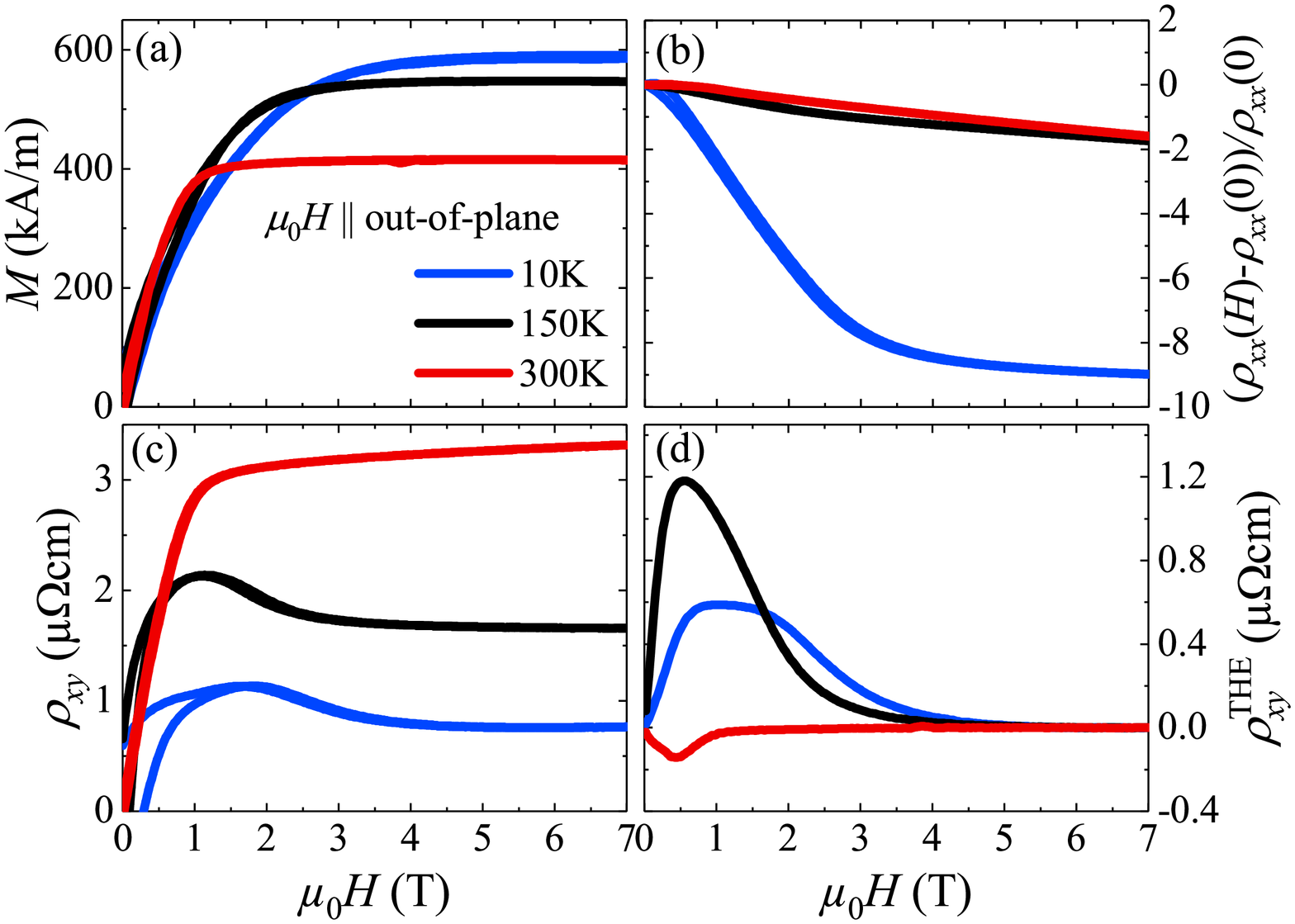}
    \caption{1st quadrant of magnetization curve and resistivities at 10~K, 150~K, and 300~K as a function of applied field in the 104~nm thick Mn$_{1.5}$PtSn film. (a) Out-of-plane magnetization, (b) magnetoresistance ratio ($H$ oriented out-of-plane), (c) Hall resistivity (Eq.~\ref{eq:Hall}) and (d) topological Hall resistivity after subtraction (Eq.~\ref{eq:fit_THE}) with averaged sweep directions.}
    \label{fig:the}
\end{figure}

The out-of-plane magnetization $M$ for the 104~nm thick Mn$_{1.5}$PtSn film is shown for 10~K, 150~K, and 300~K at magnetic fields up to 7~T in Fig.~\ref{fig:the}(a). The $M(H)$ loops are reminiscent of hard-axis behavior with a small coercive field. We attribute this to the tetragonal $c$ axis lying in the film plane. Here, the saturation magnetization $M_{s}$ is 415~kA/m, 550~kA/m and 590~kA/m at 300~K, 150~K, and 10~K, respectively, which is comparable to $M_{s}$ determined for the bulk material~\cite{Nayak2017}. The saturation field is estimated to be about 1.2~T at 300~K, increasing to about 3.5~T at 10~K. 

The magneto-resistance (MR) in Fig.~\ref{fig:the}(b), recorded with the $H$ applied along the out-of-plane direction, is depicted as the ratio $(\rho_{xx}(H)-\rho_{xx}(0))/\rho_{xx}(0)$. The MR is negative for all temperatures and is composed of two parts: First, a steep part leveling off around 4~T and visible at 10~K. This likely originates from the alignment of the spins in the non-coplanar phase and scales with the magnetometry data (Fig.~\ref{fig:the}(a)). Second, a linear field dependent part which does not saturate at 7~T. Furthermore, the absolute value of the MR ratio at 7~T clearly decreases with increasing temperature.

The Hall resistivity at 300~K in Fig.~\ref{fig:the}(c) resembles $M(H)$ (Fig.~\ref{fig:the}(a)) with a steep increase at low fields and and a linear behavior at high fields. Those two regimes can be attributed to the AHE and the OHE, respectively. Below $T_{s}$, at 150~K and 10~K, an additional non-linear part appears up to approximately 4~T. Here, $\rho_{xy}$ does not trace $M(H)$, which is reminiscent of the THE. The three different contributions can be summarized as:
\begin{equation}\label{eq:Hall}
\rho_{xy}=\rho_{xy}^{\mathrm{OHE}}+\rho_{xy}^{\mathrm{AHE}}+\rho_{xy}^{\mathrm{THE},}
\end{equation}
where $\rho^{\mathrm{OHE}}$ corresponds to the OHE scaling linearly with applied field ($\mu_{0}$H), \ahe is the AHE scaling with the magnetization component perpendicular to the film and $\rho^{\mathrm{THE}}$ represents the THE.

The AHE can arise from intrinsic and/or extrinsic mechanisms scaling with different powers of the resistivity~\cite{Nagaosa2010}. Therefor, we write $\rho^{\mathrm{AHE}}=(S_{\mathrm{A}}\rho_{xx}^{2}+\alpha\rho_{xx})M$, with $S_{\mathrm{A}}$ corresponding to intrinsic and side-jump scattering and $\alpha$ corresponding to skew scattering. In an independent analysis, we determined from the scaling relation $\rho_{xy}\propto \rho_{xx}^{\beta}$ that the underlying mechanism is of primarily of intrinsic origin with $\beta = 2.2$ (see Fig.~S3 in the Supplemental Material~\cite{Supp}). The zero-field conductivity $\sigma_{xx}\approx10^{-4}Sm^{-1}$, supports the notion that the intrinsic and side jump mechanisms dominate~\cite{Miyasato2007,Nagaosa2010}. Therefore, we focus on the skew scattering independent contributions in our evaluation in the following, taking $\alpha=0$.

In order to quantify the different contributions to the field dependant Hall resistivity, we follow the customarily performed separation process~\cite{Kanazawa2011}. Therefore, we take into account that only the AHE and OHE contribute to the Hall resistivity once the magnetization is saturated at high fields. Hence, $R_{0}$ and $S_{\mathrm{A}}$ can be obtained through a linear fit to our transport data taken at high magnetic fields, using the resistivity $\rho_{xx}$ and the (separately measured) magnetization as $\rho_{xy}/H=R_{0}+S_{\mathrm{A}}\rho_{xx}^{2}M/H$. Finally, we can calculate the topological Hall resistivity as 
\begin{equation}\label{eq:fit_THE}
\rho_{xy}^{\mathrm{THE}}=\rho_{xy}-R_{0}H+S_{\mathrm{A}}\rho_{xx}^{2}M.
\end{equation}

\begin{figure}[tp]
    \centering
    \includegraphics[width=0.9\columnwidth]{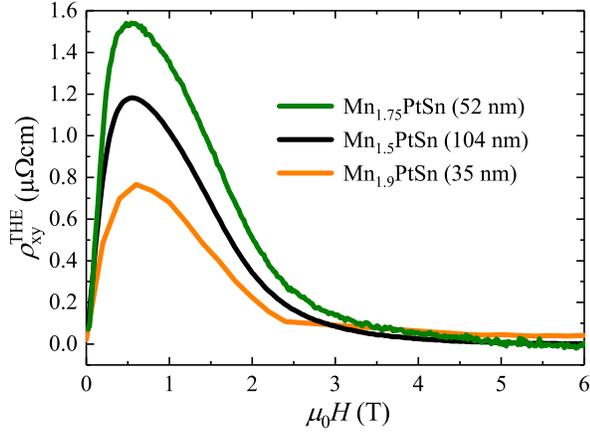}
    \caption{Topological Hall resistivity (after subtraction with averaged sweep directions) at 150~K as a function of applied field for Mn\textsubscript{1.5}PtSn (104~nm), Mn\textsubscript{1.75}PtSn (52~nm) and Mn\textsubscript{1.9}PtSn (35~nm).}
    \label{fig:the_2}
\end{figure}

As evident from Fig.~\ref{fig:the}(d) the THE in the 104~nm thick Mn$_{1.5}$PtSn film can be observed up to fields of $\mu_{0}H \approx 4$~T with a maximum topological Hall resistivity \themax $=1.2~\mu\Omega$cm at 150~K. From an anolagous analysis in films of Mn$_{1.75}$PtSn (52~nm) and Mn$_{1.9}$PtSn (35~nm) we obtained \themax $=1.55~\mu\Omega$cm and \themax $=0.76~\mu\Omega$cm, respectively at 150~K (Fig.~\ref{fig:the_2}). Our data shows that a large THE is present in a wide range of stoichiometries, underlining the robustness of the effect. This is in agreement with the presence of a (weaker) THE, previously reported in bulk \mpsa\cite{Liu2018} and polycrystalline \mpsa films ($I\overline{4}m2$)~\cite{Li2018}. Notably, in single crystalline \mpsa films ($I\overline{4}m2$)~\cite{Jin2018} with the $c$ axis in the plane, no $T_{s}$ and no THE were observed. We therefore propose that the contradicting observations (presence or absence of the THE in seemingly similar thin films) might be attributed to the different crystal structures and crystal orientations relative to the applied field.

\begin{figure}
    \centering
    \includegraphics[width=\columnwidth]{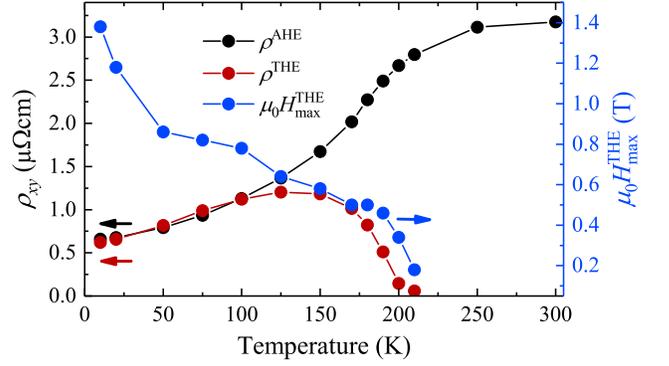}
    \caption{Anomalous and topological Hall resistivities and field of maximal topological Hall effect as a function of temperature in 104~nm thick Mn$_{1.5}$PtSn film. The anomalous Hall resistivity was obtained by extrapolating the linear high-field part of the Hall resistivity to 0~T. The maximal topological Hall resistivity is the peak of the non-linear curvature for each respective temperature at $\mu _{0}H_{\mathrm{max}}^{\mathrm{THE}}$.}
    \label{fig:poi}
\end{figure}

Figure~\ref{fig:poi} summarizes the evolution of $\rho^{\textrm{AHE}}$, \themax and the field $\mu_{0}H^{\mathrm{THE}}_{\mathrm{max}}$ at which THE reaches its maximum, with temperature. \ahe decreases continuously with temperature, having the largest slope around $T_{s}=185$~K. The THE appears below $T_{s}$, and thus must be connected with a non-coplanar spin texture at finite fields, with \themax peaking at 150~K. Interestingly, \themax and \ahe have the same magnitude between 100~K and 10~K, suggesting that a similar microscopic mechanisms is responsible for both effects. The field at which the maximal topological Hall resistivity is observed increases continuously with decreasing temperature following the same trend as the saturation field in the magnetization (Fig.~\ref{fig:the}(a)). 

Since magnetization experiments in films are challenging, the employed extraction procedure is highly susceptible to small misalignments in sample mounting or temperature differences between the transport and magnetometry measurements. This can result in significant errors of the THE values or even mimic non-existent effects. It is therefore unclear whether the THE signature at low fields and above $T_{s}$ (Fig.~\ref{fig:the}(d)) is genuine or attributable to the THE extraction process~\cite{Ishizuka2018}. However, our findings would agree with the presence of antiskyrmions above $T_{s}$ in Mn\textsubscript{1.4}Pt\textsubscript{0.9}Pd\textsubscript{0.1}Sn~\cite{Nayak2017}.

In contrast to the majority of reports on the THE in conjunction with the AHE, we find that in Mn$_{1.5}$PtSn thin films the size of the THE is of the same magnitude as the corresponding AHE. Typically, the AHE by far surpasses the THE~\cite{NeuBauer2009,Huang2012,Rana2016,Spencer2018}. Nevertheless, similar behavior as in Mn$_{1.5}$PtSn was also presented in the non-collinear metallic Mn\textsubscript{5}Si\textsubscript{3} and the correlated oxide charge-transfer insulator (Ca,Ce)MnO$_{3}$~\cite{Suergers2014,Vistoli2018}. Interestingly, one can also find a few examples where the THE appears in conjunction with a vanishing AHE, such as the Weyl semimetal GdPtBi and the helimagnetic metal MnGe~\cite{Suzuki2016,Kanazawa2011}. Thus, the dependence of the underlying mechanism (i.e. skyrmions/bubbles, Weyl points or non-coplanar magnetic structure) in the respective material system (e.g. thin film, bulk or multilayer) determines the relation of the THE to the AHE, which can range over several orders of magnitude. The physics regarding the relation of the THE to the AHE have not been completely explored or understood, where in our films we clearly observe a difference in the relation that depends on the spin reorientation transition temperature. 

%%%%%%%%%%%%%%%%%%%%%%%%%%%%%%%%%%%%%%%%%%%%%%%%%%%%%%%%%%%%%

\section{\label{sec:level4}Conclusion}
In this work, we report a non-trivial behavior of the Hall response in \mpsb thin films (space group $I\overline{4}2d$) identified as the THE. The signature is clearly evident even prior to the customarily performed subtraction of magnetometry data. The THE is present up to a spin reorientation transition temperature, $T_{s}=185$~K, and a field of $\mu_{0}H \approx 4$~T. The same magnitude of \themax and \ahe below 100~K implies a similar microscopic mechanism for the AHE and THE. While we focused on a 104~nm thick Mn$_{1.5}$PtSn film, similar experiments in different Mn$_{2-x}$PtSn films show that the THE is robust over various stoichiometries and thicknesses, reaching up to \themax $=1.55~\mu\Omega$cm at 150~K. All together, Mn$_{2-x}$PtSn is an interesting compound for the understanding and application of transport phenomena in topological magnetic structures.

%%%%%%%%%%%%%%%%%%%%%%%%%%%%%%%%%%%%%%%%%%%%%%%%%%%%%%%%%%%%%

\section*{acknowledgement}
The authors acknowledge funding by the Deutsche Forschungsgemeinschaft (DFG, German Research Foundation) under SPP 2137 (Project number 403502666),  ERC Advanced Grant 742068 “TOPMAT” and EU FET Open RIA Grant No. 766566 grant (ASPIN). P. S. acknowledges financial support by the International Max Planck Research School for Chemistry and Physics of Quantum Materials (IMPRS-CPQM). 

%%%%%%%%%%%%%%%%%%%%%%%%%%%%%%%%%%%%%%%%%%%%%%%%%%%%%%%%%%%%%


%%%%%%%%%%%%%%%%%%%%%%%%%%%%%%%%%%%%%%%%%%%%%%%%%%%%%%%%%%%%%%

\end{document}